\documentstyle[twocolumn,aps,prd]{revtex}

\begin{document}
\input{psfig}
\draft
\title{High energy parton-parton amplitudes from lattice 
QCD and the stochastic vacuum model}  
\author{A. F.  Martini, M. J.  Menon and D. S. Thober}
\address{Instituto de F\'{\i}sica Gleb Wataghin, \\ 
Universidade Estadual de Campinas, Unicamp, \\
13083-970 Campinas, SP, Brazil} 
\maketitle
\begin{abstract}
Making use of the gluon gauge-invariant two-point
correlation function, recently determined by numerical
simulation on the lattice in the quenched approximation and 
the stochastic vacuum model, 
we calculate the elementary (parton-parton) amplitudes 
in both impact-parameter and momentum transfer spaces. 
The results are compared with those obtained 
from the Kr\"{a}mer and Dosch ansatz for the correlators. Our 
main conclusion is that the divergences in the correlations 
functions suggested by the lattice calculations do not affect 
substantially the elementary amplitudes. 
Phenomenological and semiempirical information presently 
available on elementary amplitudes is also referred to and is 
critically discussed in connection with some theoretical issues.
\end{abstract}

 
\section{Introduction}
\label{sec:intro}

The theoretical investigation of quark-quark scattering is one of 
the topical problems in high-energy hadron physics. In principle, 
the possibility to predict this amplitude directly from quantum 
chromodynamics (QCD) could lead to important insight concerning 
high-energy soft processes and meanly elastic hadron scattering. 
Although some phenomenological and semiempirical information 
already exists \cite{menon93,furget90}, the determination of this 
amplitude from a pure QCD approach (model independent) is still 
an open problem. However, recently, remarkable progress has been 
achieved, starting with the nonperturbative approach by Landshoff 
and Nachtmann \cite{landshoff87}, in which quarks are coupled to 
Abelian gluons (gluon condensate). In a non-Abelian model, the 
amplitudes of quarks at high energies were computed by Nachtmann 
through the use of the eikonal method \cite{nachtmann91}. 
Analogous version was then developed by Kr\"{a}mer and Dosch 
\cite{kramer90} in the context of the stochastic vacuum model 
(SVM) \cite{dosch87,dosch88}. This model describes the low 
frequencies contributions in the functional integral of QCD 
in terms of a stochastic process by means of a cluster expansion 
\cite{kampen74} and gives a good description of the heavy quark 
potential \cite{simonov89}. The most general form 
of the lowest cluster is the gauge invariant two-point field 
strength correlator \cite{dosch87,dosch88,dosch94}
\newpage

\begin{eqnarray} 
& &\langle{\bf{F}}_{\mu \nu}^{C}(x){\bf{F}}_{\rho
\sigma}^{D}(y)\rangle \nonumber\\
& &={\delta}^{CD}g^{2}\frac{\langle FF\rangle}{12(N_c^2-1)}\Biglb(
({\delta}_{\mu\rho}{\delta}_{\nu \sigma}-{\delta}_{\mu \sigma}
{\delta}_{\nu\rho}){\kappa}D({z^2/a}^{2}) \nonumber \\
& &+\frac{1}{2}[{\partial}_{\mu}(z_{\rho}{\delta}_{\nu
\sigma}-z_{\sigma}{\delta}_{\nu
\rho})+{\partial}_{\nu}(z_{\sigma}{\delta}_{\mu
\rho}-z_{\rho}{\delta}_{\mu \sigma})]\nonumber\\
& &\times (1-{\kappa})D_{1}({z^2/a}^{2})\Bigr),
\label{gluoncorrel}
\end{eqnarray} 
where $z=x-y$ is the two-point distance, $a$ is a characteristic 
correlation length, ${\kappa}$ a constant,  $g^{2}\langle FF\rangle$ 
the gluon condensate, and $N_c$ the number of colors, $C, 
D=1,...,N_c^2-1$ . The two scalar functions $D$ and $D_{1}$ describe 
the correlations and are normalized as $D(0)=D_{1}(0)=1$.

These correlation functions play a central role in the application 
of the SVM to high-energy scattering. Once one has information about 
$D$ and $D_{1}$, the SVM leads to the determination of the elementary 
quark-quark scattering amplitude and this is the point we are 
interested in.

Numerical determinations of these correlation functions, in limited 
interval of physical distances, exist from lattice QCD in the 
quenched approximation. The first 
determination, by Di Giacomo and Panagopoulos, covered the interval 
0.4 to 1.0 fm \cite{giacomo92}. Recently, through improved technique 
(larger lattice) the range was extended down to 0.1 fm by Di Giacomo, 
Meggiolaro, and Panagopoulos \cite{giacomo97} and this brought new and 
important insights on the subject, as will be discussed.

On the other hand, Kr\"{a}mer and Dosch introduced a suitable ansatz 
for the correlation function $D$ and $D_1$\cite{kramer90}, which is 
in agreement with the early lattice results in the range 0.4-1.0 fm 
\cite{dosch94}. Using the SVM they acchieved good descriptions of the 
experimental data on total cross sections and slopes of the elastic 
amplitudes  in hadronic processes \cite{kramer90,dosch94}. Also, with 
the same ansatz, Grandel and Weise calculated differential cross 
sections through the eikonal approximation (multiple diffraction 
Glauber theory). Introducting a monopole parametrization for the form 
factors, a satisfactory description of experimental data on $pp$ and 
$\overline{p}p$ elastic scattering at small momentum transfer was 
obtained \cite{grandel95}.

In this paper we make use of the recent lattice results for the 
correlation functions and, through the SVM, we calculate the gauge 
invariant elementary amplitudes in both impact-parameter and 
momentum transfer spaces. The results are then compared with those 
obtained through the Kr\"{a}mer-Dosch ansatz and also with some 
phenomenological and semiempirical information available. 
Since in the SVM the amplitudes are characterized by scattering 
amplitudes for Wilson loops in Minkowski space, they will 
generically be referred to as elementary {\it parton-parton} 
amplitudes.

The material is organized as follows. In Sec. 
\ref{sec:partoamplsvm} we briefly recall the essential formulas 
of the SVM connecting the field strength correlator 
(\ref{gluoncorrel}) with the parton-parton amplitudes and in Sec. 
\ref{sec:correfunc} we review the correlation functions from 
lattice calculations and the ansatz by Kr\"{a}mer and Dosch. 
With this information in Sec. \ref{sec:partoampl} we present 
our calculations and results in some detail. Discussions and 
critical remarks concerning nonperturbative QCD, phenomenological, 
and semiempirical results on parton-parton amplitudes are 
presented in Sec. \ref{sec:discu} and conclusions in Sec. 
\ref{sec:finconcl}.

\section{Parton-parton amplitudes in the stochastic vacuum 
model}
\label{sec:partoamplsvm}

In the nonperturbative QCD framework referred to in the last 
section, the study of the elementary scattering is based on the 
amplitudes of quarks moving on lightlike paths in an external 
field. In the Nachtmann approach the quarks involved in a 
scattering pick up an eikonal phase in traveling through the 
nonperturbative QCD vacuum. In order to have gauge invariant 
Dirac wave function solutions a Wilson loop is proposed to 
represent each quark. In this context the no-color exchange 
parton-parton (loop-loop) amplitude can be written as 
\cite{nachtmann91}

\begin{eqnarray}
{\gamma}&=&{\langle}{\rm Tr}[{\cal{P}}e^{-ig{\int}_{\rm loop 1}
d{\sigma}_{\mu\nu}F_{\mu \nu}(x;w)}-1]\nonumber \\
&\times&{\rm Tr}[{\cal{P}}
e^{-ig{\int}_{\rm loop2}d{\sigma}_{\rho\sigma}F_{\rho \sigma}(y;w)}
-1]{\rangle}, 
\end{eqnarray}
where ${\langle}{\rangle}$ means the functional integration over the 
gluon fields, the integrations are over the respective loop areas, 
and $w$ is a common reference point from which the integrations are 
performed.

This expression is simplified in the Kr\"{a}mer and Dosch 
description by taking the Wilson loops on the light-cone. In the 
SVM the leading order contribution to the amplitude is given by 
\cite{kramer90}

\begin{equation}
 {\gamma}(b)=\eta{\epsilon}^{2}(b) ,
\label{perfilepsi}
\end{equation}
where $b$ is the impact parameter, $\eta$ is a constant depending 
on normalizations [see Eqs.\ (\ref{fator98giaco}) and 
(\ref{fator98dosch})], and
\begin{equation}
 {\epsilon}(b)=g^{2}{\int}{\int}d{\sigma}_{\mu
\nu}d{\sigma}_{\rho \sigma}{\rm Tr}{\langle}F_{\mu \nu}(x;w)F_{\rho
\sigma}(y;w){\rangle}. 
\end{equation} 
Here ${\langle}g^2F_{\mu \nu}(x;w)F_{\rho\sigma}(y;w){\rangle}$ is 
the Minkowski version of the gluon correlator. We will return to 
this point later.

After a two-dimensional integration, $\epsilon(b)$ may be 
expressed in terms of the correlation functions in 
Eq.\ (\ref{gluoncorrel}) by \cite{kramer90}

\begin{eqnarray}
\epsilon(b)=\epsilon_{\rm I}(b)+\epsilon_{\rm II}(b),
\label{epsilont}
\\
\epsilon_{\rm I}(b)={\kappa}{\langle}g^2FF{\rangle}
{\int}_{b}^{\infty}db'(b'-b){\cal{F}}_{2}^{-1}
[D(-q^2)](b')
\label{epsilonI}
\\
\epsilon_{\rm II}(b)=({1-\kappa}){\langle}g^2FF{\rangle}
{\cal{F}}_{2}^{-1}\left[\frac{d}{dq^{2}}D_{1}(-q^2)\right](b),
\label{epsilonII}
\end{eqnarray}
where $q^2$ is the four-momentum transfer squared and 
${\cal{D}}=D$ or $D_1$,

\begin{equation}
{\cal{D}}(k^2)={\cal{F}}_4[{\cal{D}}(z^2)]
\label{transf4d}
\end{equation}
with ${\cal{F}}_n$ denoting a $n$-dimensional Fourier transform.

In the impact parameter space $\gamma$ represents the elementary 
profile function, from which the elementary parton-parton 
scattering amplitude is calculated through a two-dimensional 
Fourier transform:
\begin{equation}
f(q^2)=\int_0^{\infty}bdb J_0(qb)\gamma(b) ,
\label{amplperfil}
\end{equation}
where $J_0$ is the Bessel function.

With the above formalism, once one has inputs for the correlation 
functions $D(z)$ and $D_1(z)$ the elementary amplitudes in the 
impact parameter and transfer momentum spaces, Eqs. 
(\ref{perfilepsi}) and (\ref{amplperfil}), respectively, may, in 
principle, be evaluated through Eqs. 
(\ref{epsilont})-(\ref{transf4d}).

\section{Correlation functions}
\label{sec:correfunc}

We now recall the theoretical information available about the 
correlation functions $D$ and $D_1$, namely, that obtained from 
lattice QCD and the ansatz introduced by Kr\"{a}mer and Dosch. 
In Sec. \ref{sec:partoampl} we use these functions in order to 
calculate the elementary amplitudes. The translation from Euclidean 
to Minkowski space will be discussed in the next section.

\subsection{Lattice results}
\label{subsec:lattiresu}

The first determination of the correlation functions from lattice 
QCD in the quenched approximation was obtained through the cooling 
technique \cite{campostrini89,giacomo90}, with a lattice of size 
$16^4$ and in the interval of physical distance 0.4 to 1.0 fm 
\cite{giacomo92}. In this range the theoretical data showed 
agreement with an exponential decrease of both correlators with the 
distance. In particular, using the notation of Ref. \cite{dosch94}, 
the function $D$ is parametrized by 

\begin{equation}
{\kappa}{\langle}g^2FF{\rangle}D\Bigl(-{r^2\over a^2}\Bigr)=24C
\exp{\Bigl(-{r\over \lambda}\Bigr)},
\label{Ddedosch}
\end{equation}
where $r$ is the physical distance (Euclidean space), and 

\begin{equation}
\lambda={1\over 183\Lambda},\qquad {C\over \Lambda^4}=3.6\times 
10^8,
\label{lambdacdedosch}
\end{equation}
with a typical statistical error of a few percent. 
In Ref. \cite{giacomo92} the value of $\Lambda$ was determined 
from string tension studies \cite{michael88} leading to

\begin{equation}
C=137\; {\rm fm}^{-4},\qquad \lambda=0.22\; {\rm fm},
\label{alambdadedosch}
\end{equation}
which corresponds to $\Lambda = 4.9$ MeV.

Recently new results were obtained on a $32^4$ lattice allowing the 
determination of the correlators at distances down to 0.1 fm 
\cite{giacomo97}. A novel result was the appearance of a deviation 
from the exponential behavior at shorter distances ($< 0.4$ fm), 
indicating a $1/x^4$ divergence at the origin. Putting together the 
data of Refs. \cite{giacomo92} and \cite{giacomo97}, good agreement 
is obtained within statistical errors, with a parametrization of the 
form \cite{giacomo97}

\begin{eqnarray}
& &{\kappa}{\langle}g^2FF{\rangle}D(z^2)\nonumber\\
& &=24\left[A\exp\Bigl(-{|z|\over 
\lambda_A}\Bigr)+{B\over |z|^4}\exp\Bigl(-{|z|\over \lambda_B}
\Bigr)\right],
\label{Ddegiacomo}
\\
& &(1-{\kappa}){\langle}g^2FF{\rangle}D_1\nonumber\\
& &=24\left[A_1
\exp\Bigl(-{|z|\over \lambda_A}\Bigr)+{B_1\over |z|^4}
\exp\Bigl(-{|z|\over \lambda_B}\Bigr)\right].
\label{D1degiacomo}
\end{eqnarray}
As before, from string tension information on $\Lambda$, 
$\lambda_A=0.22$ fm and $\lambda_B=0.43$ fm. The typical error 
is again of the order of a few percent. 
In Sec. \ref{sec:partoampl} we test and discuss the influence of 
the divergent terms.

\subsection{Kr\"{a}mer-Dosch ansatz}
\label{subsec:kramedosch}

In the approach of Ref. \cite{dosch94}, the effect of 
$D_1(z^2/a^2)$ is neglected since, 
from QCD lattice calculation, the value of $\kappa$ in 
Eq. (\ref{gluoncorrel}) is $\approx 1$ \cite{giacomo92}. So, we will 
consider only $D$ for this case.

Assuming functions which have a well defined Fourier transform 
and that can be analitically continuated to Euclidean world in the 
momentum space, Kr\"{a}mer and Dosch introduced a family of functions 
depending on an integer parameter $n$ \cite{kramer90}. 
These functions may be analitically integrated and the selection 
of the $n$ value was made by comparison with the early lattice 
results, Eq. (\ref{Ddedosch}). In the Euclidean space the fit in the 
interval 0.5--0.8 fm (early lattice results) led to $n=4$ and for 
$\Lambda=4.4\;$ MeV it was obtained $a=0.35$ fm and 
${\kappa}{\langle}g^2FF{\rangle}=1.774\;{\rm GeV}^4$ \cite{dosch94}. 
The final result for the correlation function with $n=4$ is given by 
\cite{dosch94}

\begin{equation}
D_{\rm KD}^{(4)}(x)=x\left[K_1(x)-{x\over 4}K_0(x)\right],
\label{D4dedosch}
\end{equation}
where $K_1,K_0$ are Bessel functions and $x$ is a dimensionless 
variable
\[ x={3\pi\over 8}{z\over a}. \]

\subsection{Remarks on the correlation functions}
\label{subsec:remacorr}
In what follows, it is important to stress that the 
Kr\"{a}mer-Dosch ansatz for the correlation function $D$ is an 
analytical form, firstly selected by analytical continuation and 
asymptotic limit conditions and then through fit to early lattice 
information, Eq. (\ref{Ddedosch}). On the other hand, ``real'' 
lattice results correspond to sets of discret theoretical points 
with errors, in a finite interval of physical distances 
(0.1 -- 1.0 fm). The parametrizations introduced in both early 
\cite{giacomo92} and recent \cite{giacomo97} 
results, extrapolate this interval down and above. Although, 
presently, there is no theoretical informations on these 
extrapolated regions we will assume the parametrizations 
(\ref{Ddegiacomo}) and (\ref{D1degiacomo}) as representing lattice 
results and will refer as {\it lattice parametrizations}. We will 
return to this point in Secs. \ref{subsec:resulatt} and 
\ref{subsec:theolimi}.

\begin{figure}[h]
\centerline{
\psfig{figure=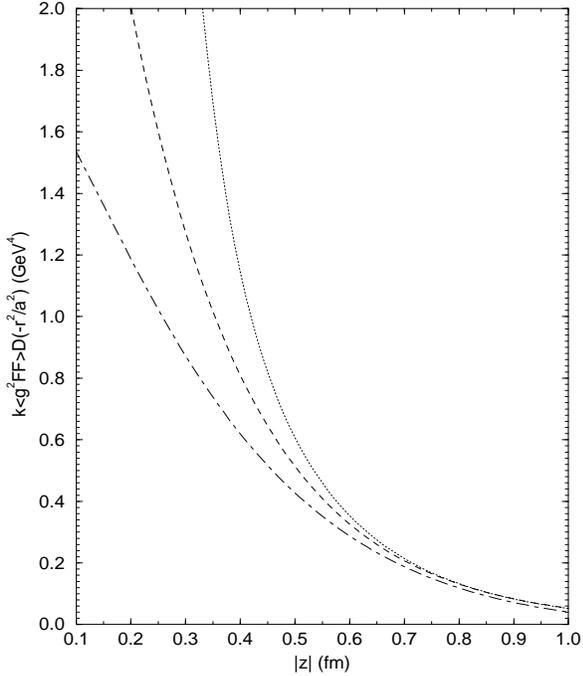,height=8.0cm,width=8cm}}
\vspace*{10.0mm}
\caption{Correlation function $D$ given by (a) recent lattice 
parametrization, Eq. (\protect\ref{Ddegiacomo}) (dotted); (b) early 
lattice parametrization, Eq. (\protect\ref{Ddedosch}) (dashed); 
(c) Kr\"{a}mer and Dosch ansatz, Eq. (\protect\ref{D4dedosch}) 
(dot-dashed).}
\label{correldosch}
\end{figure}

The results reviewed in this section for the correlation function 
$D$ are displayed in Fig.\ \ref{correldosch} for comparison and this 
deserves some discussion. We plotted the curves corresponding to 
the early lattice result, Eqs. (\ref{Ddedosch}) and 
(\ref{alambdadedosch}), the new lattice result, 
Eq. (\ref{Ddegiacomo}), and the Kramer-Dosch ansatz, 
Eq. (\ref{D4dedosch}). As mentioned above, the 
final form of this ansatz was obtained through a fit to the early 
lattice parametrization in the interval $0.4 - 0.8$ fm. From 
Fig.\ \ref{correldosch}, there is a discrepancy between these 
two curves in the above interval, which does not appear in 
Fig. 10 of Ref. \cite{dosch94}. The reason is the different values 
of the parameter $\Lambda$, namely, 4.4 MeV in Ref. \cite{dosch94} 
and 4.9 MeV in Ref. \cite{giacomo92} and Fig.\ \ref{correldosch}. 
This difference is significant since, for example, the parameter 
$C$ in Eq. (\ref{lambdacdedosch}) is proportional to $\Lambda^4$. 
On the other hand, the differences between early and new lattice 
results in the above interval are compatible with the errors 
associated with each curve.

\section{Parton-parton amplitudes}
\label{sec:partoampl}

In the last section we reviewed the theoretical information 
presently available concerning the correlation functions, namely, 
the recent lattice parametrizations, Eqs. (\ref{Ddegiacomo}) and 
(\ref{D1degiacomo}), and the result from the ansatz by Kr\"{a}mer 
and Dosch, Eq. (\ref{D4dedosch}).

In the context of the SVM (Sec. \ref{sec:partoamplsvm}), the 
corresponding predictions for the parton-parton amplitudes are 
obtained, in principle, through four steps, envolving the 
following calculations: (1) $D(k)$ and $D_1(k)$ by the 
four-dimensional Fourier transform, Eq. (\ref{transf4d}); 
(2) two-dimensional inverse Fourier transforms of $D(k)$ and 
$dD_1/dk^2$, which enter in 
Eqs.\ (\ref{epsilonI}) and (\ref{epsilonII}); 
(3) eikonal phase through Eqs. (\ref{epsilonI}) and 
(\ref{epsilonII});
(4) elementary amplitudes in the impact parameter space 
(profile function), Eq. (\ref{perfilepsi}), and in the transfer 
momentum space (scattering amplitude), Eq. (\ref{amplperfil}). 
These are the central points of this work and in this section we 
present and discuss the calculations in some detail.
\newpage

\subsection{Results from lattice parametrizations}
\label{subsec:resulatt}

Our basic assumptions in using lattice results are the following.

(a) We consider the parametrizations (\ref{Ddegiacomo}) 
and (\ref{D1degiacomo}) as representing the correlation functions 
and so, the correlation lenght has the value $a=0.22$ fm.

(b) The effect of the extrapolation down to 0.1 fm will 
be investigated by taking account or not of the divergent term 
$1/|z|^4$, that is, taking or not $B=0$ and $B_1=0$ in Eqs. 
(\ref{Ddegiacomo}) and (\ref{D1degiacomo}), respectivelly.

(c) We consider the lattice parametrizations, Eqs. 
(\ref{Ddegiacomo}) and (\ref{D1degiacomo}), as {\it representing} 
the correlator functions in the Euclidean world. With this 
assumption the above correlation functions enter into 
Eq. (\ref{gluoncorrel}), with the adequated tensor structures, 
and then directly in Eq. (\ref{transf4d}).

In using the above assumptions we seek to see the consequences in 
terms of the elementary amplitudes.

\begin{figure}
\centerline{
\psfig{figure=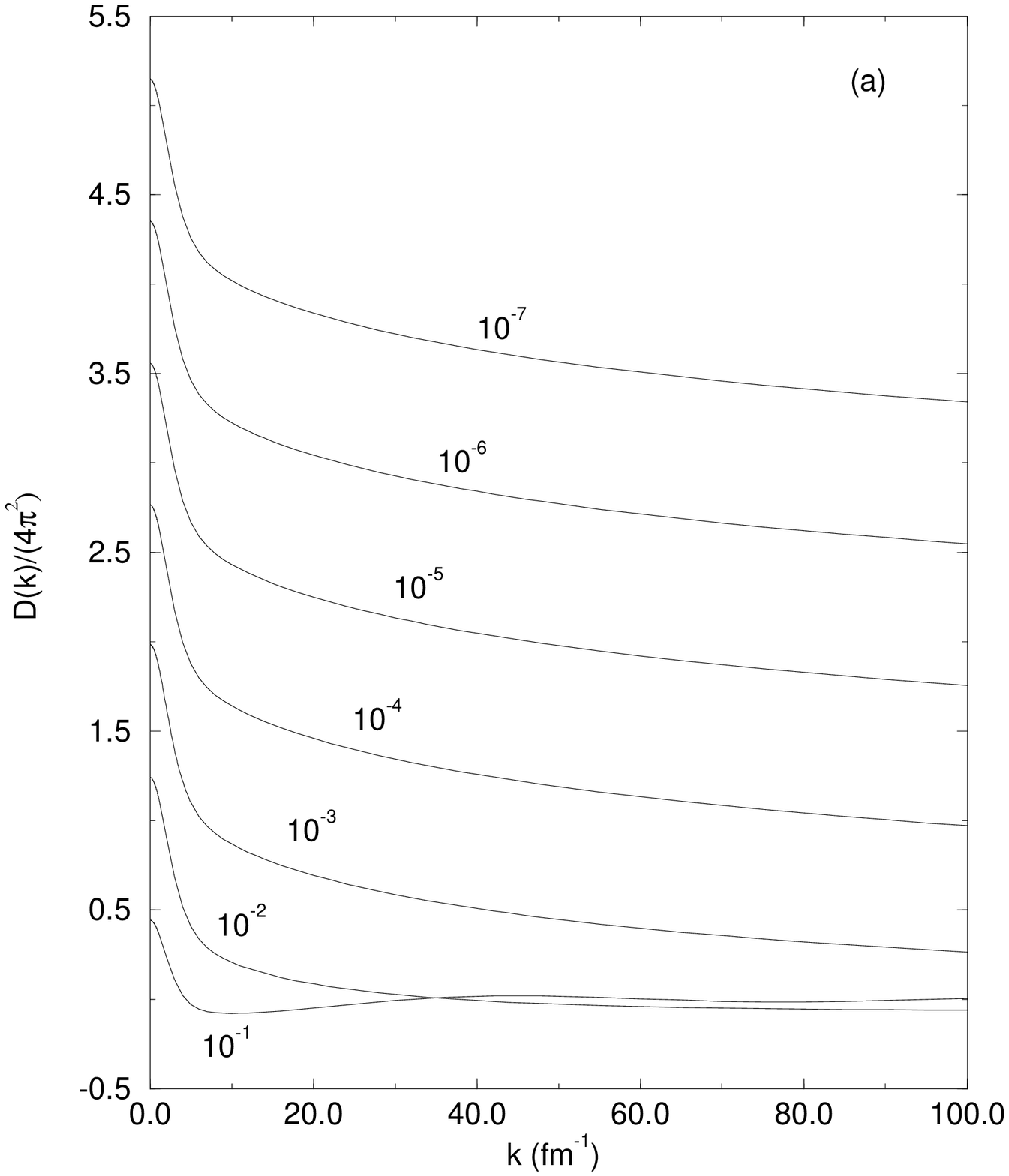,height=8.0cm,width=8cm}}
\vspace*{10.0mm}
\centerline{
\psfig{figure=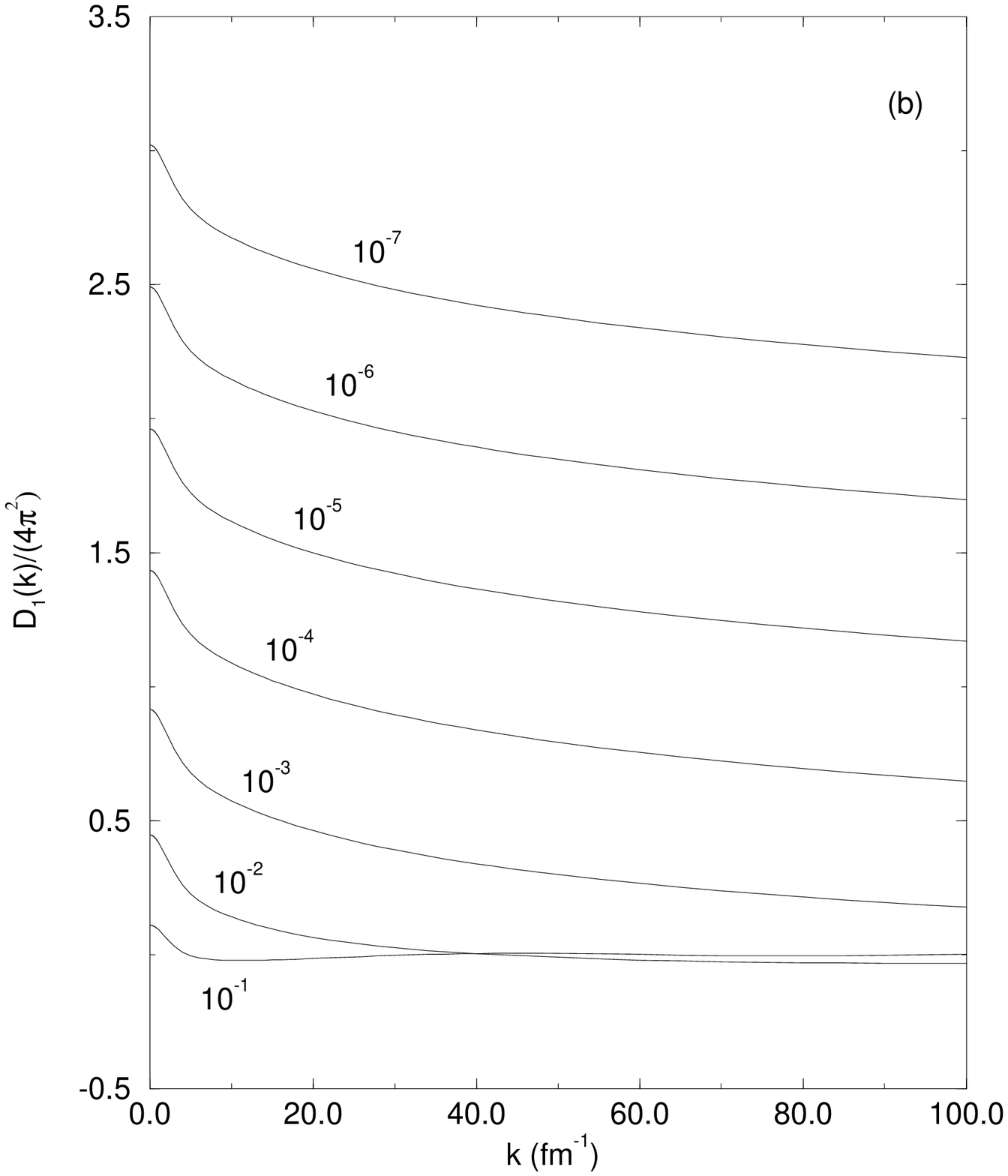,height=8.0cm,width=8cm}}
\vspace*{10.0mm}
\caption{Four-dimensional Fourier transform of $D(z)$, Eq. 
(\protect\ref{transf4dDz}), and the effect of the lower integral 
limit $z_m$ (between $10^{-1}$ and $10^{-7}$) as represented by 
Eq. (\protect\ref{Dmaisconst}): (a) $\hat{d}(k)=d(k)$; (b) $
\hat{d}(k)=d_1(k)$.}
\label{dd1k101a7}
\end{figure}

The first step referred to before concerns the four-dimensional 
transform 

\begin{equation}
{\cal{D}}(k)=F_4[{\cal{D}}(z)]=\int d^4z{\cal{D}}(z)
\exp(i{\bf k}\cdot {\bf z})
\label{transf4dDz}
\end{equation}
for ${\cal{D}}=D,\;D_1$, here in Euclidean space according to 
assumption (c) above. Although the lattice data are limited to the 
interval 0.1--1.0 fm, the parametrizations (\ref{Ddegiacomo}) and 
(\ref{D1degiacomo}) extend to all the space and include the 
divergent term $|z|^{-4}$. However, we found that when the lower 
integral limit $z_m$ becomes smaller than $\approx 10^{-3}$ fm, 
numerical evaluation of this transform \cite{nag93}, for both 
correlators, may be put in the form

\begin{equation}
{\cal{D}}(k)=\hat{d}(k)+C(z_m) ,
\label{Dmaisconst}
\end{equation} 
where $\hat{d}(k)=d_1(k),\;d(k)$ are smooth (finite) decreasing 
functions of $k$, and $C(z_m)$ is a constant which increases when 
the lower integral limit becomes smaller. This effect is shown in 
Fig. \ref{dd1k101a7}.

Now, in the second step, we should 
calculate the two-dimensional Fourier transforms of $D(k)$ and 
$dD_1(k^2)/dk^2$. Because of the derivative in the last case the 
constant $C(z_m)$ may be neglected. In the former case, since 
the Fourier transform of $C(z_m)$ leads to a $\delta$ function, 
its effect in the calculation of the eikonal phase $\epsilon_I$ 
in Eq. (\ref{epsilonI}) has also no influence.

\begin{figure}
\centerline{
\psfig{figure=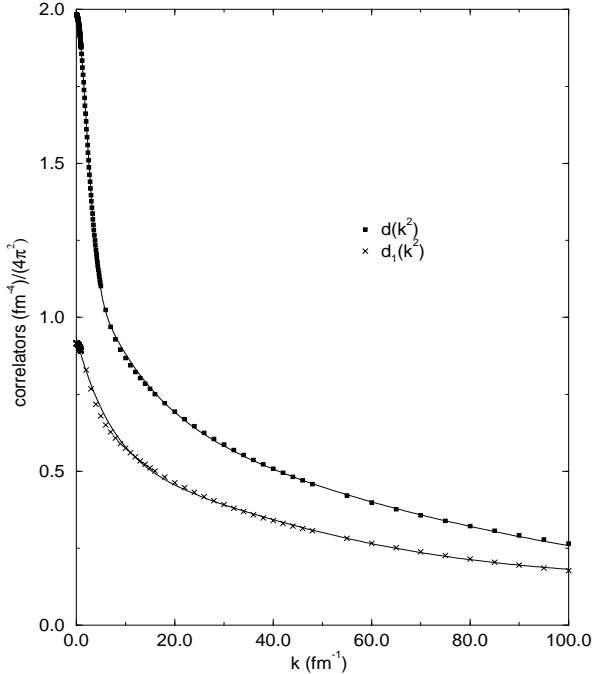,height=8.0cm,width=8cm}}
\vspace*{10.0mm}
\caption{Finite part of the correlation functions in momentum space, 
Eq. (\protect\ref{Dmaisconst}), calculated numerically (points) and 
fits through Eq. (\protect\ref{dexpogauss}) (solid curves).}
\label{dd1k}
\end{figure}

With this, carring out the numerical integration down to 
$10^{-3}$ fm we have a stable behavior for $\hat{d}(k)$ in both 
cases (see Fig. \ref{dd1k101a7}). The numerical results for 
$d(k)$ and $d_1(k)$ are shown in Fig.\ \ref{dd1k}. We then proceed 
to fit these points through the CERN-MINUIT routine \cite{james92} 
by functions of the type

\begin{eqnarray}
d(k)=\sum_{j=1}^{2}a_j\exp(-b_jk)+a_3\exp(-b_3k^2),
\nonumber
\\
d_1(k)=\sum_{j=1}^{2}a_{1j}\exp(-b_{1j}k)+a_{13}\exp(-b_{13}k^2),
\label{dexpogauss}
\end{eqnarray}
also shown in Fig.\ \ref{dd1k}. The values of the free parameters are 
displayed in Table\ \ref{paramexpogauss}.

\newpage

\begin{table}
\caption{Values of the free parameters for $d(k)$ and $d_1(k)$ in 
Eq. (\protect\ref{dexpogauss}). Parameters $b_1,\;b_2$ and 
$b_{11},\;b_{12}$ are in fm while $b_3,\;b_{13}$ in ${\rm fm}^2$.}
\label{paramexpogauss}
\begin{tabular}{ccccc}
j & $a_j/(4\pi^2)$ & $b_j$ & $a_{1j}/(4\pi^2)$ & $b_{1j}$\\
\tableline
1 & 0.50003 & 0.094168 & 0.46314 & 0.12874\\
2 & 0.76546 & 0.010859 & 0.28780 & $0.48787\times 10^{-2}$\\
3 & 0.73263 & 0.104490 & 0.18440 & $0.36088\times 10^{-3}$\\
\end{tabular}
\end{table}

With Eq.\ (\ref{dexpogauss}) and the derivative of $d_1(k)$, 
step 2 is performed analytically leading to \footnote{For short, 
in what follows, we will refer only to the analytical structure 
of the parametrized and calculated functions.}

\begin{equation}
{\cal{F}}_2^{-1}[d(-q^2)](b)=\sum_{j=1}^2\alpha_j[\beta_jb^2+
1]^{-3/2}+\alpha_3\exp(-\beta_3b^2),
\label{f2ded}
\end{equation}

\begin{eqnarray}
{\cal{F}}_2^{-1}\left[{d\over dq^2}d_1(-q^2)\right](b)&=&
\sum_{j=1}^2
\alpha_{1j}[\beta_{1j}b^2+1]^{-1/2}+\alpha_{13}\nonumber\\
& &\times\exp(-\beta_{13}b^2).
\label{f2ded1}
\end{eqnarray}

\begin{figure}
\centerline{
\psfig{figure=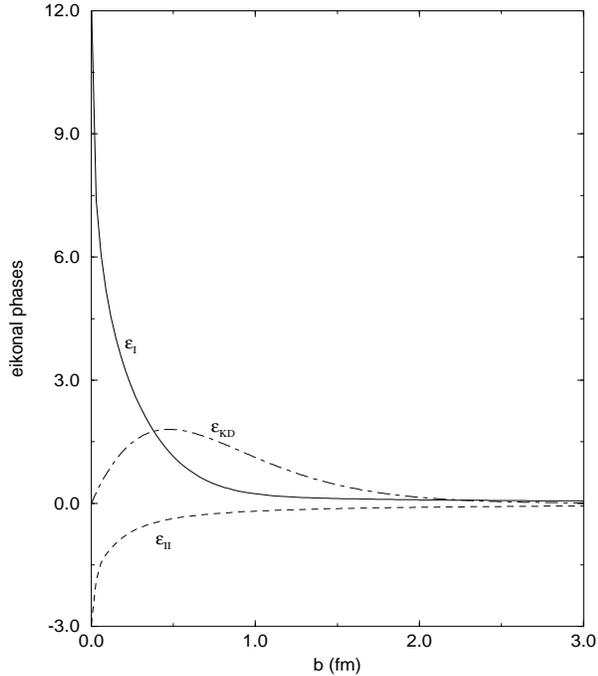,height=8.0cm,width=8cm}}
\vspace*{10.0mm}
\caption{Contributions for the eikonal phase, Eqs. 
(\protect\ref{epsilont}), (\protect\ref{epsilonI}), and 
(\protect\ref{epsilonII}), from lattice ($\protect\epsilon_{\rm I}$ 
and $\protect\epsilon_{\rm II}$ from $D$ and $D_1$, respectively) 
and Kr\"{a}mer-Dosch ansatz ($\protect\epsilon_{\rm KD}$).}
\label{epsilonIepsilonII}
\end{figure}

From Eq. (\ref{f2ded}), in step 3, the contribution 
$\epsilon_{\rm I}(b)$ to the eikonal phase is analytically 
calculated:

\begin{eqnarray}
\epsilon_{\rm I}(b)&=&\zeta_1\exp(-\xi_1b^2)+\zeta_2
[{\rm Erfc}(\xi_2b)]b+\sum_{j=1}^2a_j\Biglb([(\varepsilon_jb)^2
\nonumber\\
& &+1]^{1/2}+\varepsilon_jb\Bigrb),
\label{epsilonIErfc}
\end{eqnarray}
where Erfc is the complementary Error function. The contribution 
$\epsilon_{\rm II}(b)$, Eq. (\ref{epsilonII}), is given directly by 
Eq.\ (\ref{f2ded1}). Figure \ref{epsilonIepsilonII} shows both 
$\epsilon_{\rm I}$ and $\epsilon_{\rm II}$ as function of the impact 
parameter. We will discuss these results in Sec. \ref{sec:discu}.

At last, step 4 is performed with 
$\epsilon(b)=[\epsilon_{\rm I}(b)+\epsilon_{\rm II}(b)]$ into Eq. 
(\ref{epsilont}), leading to the elementary profile $\gamma(b)$, 
Eq. (\ref{perfilepsi}), with normalization constant 

\begin{equation}
\eta={4\over 9\times 8^2} .
\label{fator98giaco}
\end{equation}
Then numerical integration furnishes the 
corresponding elementary parton-parton amplitude $f(q)$, Eq. 
(\ref{amplperfil}). The results for $\gamma(b)$ and $f(q)$ are 
shown in Figs.\ \ref{gammab} and \ref{amplq}, respectively.

As expressed in the beginning of this section, in order to 
investigate the effect of the divergent term in the recent lattice 
results, we also calculated $\gamma(b)$ and $f(q)$ neglecting 
these terms in both correlators, i.e., by taking account only of 
the first term in Eq. (\ref{Ddegiacomo}) and (\ref{D1degiacomo}). 
All these results (with and without the divergent terms) are 
displayed in Figs.\ \ref{gammab} and \ref{amplq} for $\gamma(b)$ 
and $f(q)$, respectively. We will return to this point in 
Sec. \ref{sec:discu}, after the discussion of $\gamma(b)$ and 
$f(q)$ predicted from the Kr\"{a}mer and Dosch ansatz.
\newpage

\subsection{Results from Kr\"{a}mer-Dosch correlators}
\label{subsec:resukramdosc}

Concerning step two, a suitable parametrization for the 
two-dimensional inverse Fourier transform of $D(k)$ was 
introduced in Ref. \cite{dosch94}, so that step three may be 
performed analytically leading to 

\begin{equation}
\epsilon_{\rm KD}={\cal{K}}\exp(-x)\sum_{n=0}^4\varrho_nx^n,
\label{epsilonKD}
\end{equation}
where ${\cal{K}}={\kappa}{\langle}g^2FF{\rangle}a^42^{14}/
(3^4\pi^3)$. This result is displayed in 
Fig.\ \ref{epsilonIepsilonII} together with those obtained through 
the recent lattice results.

\begin{figure}
\centerline{
\psfig{figure=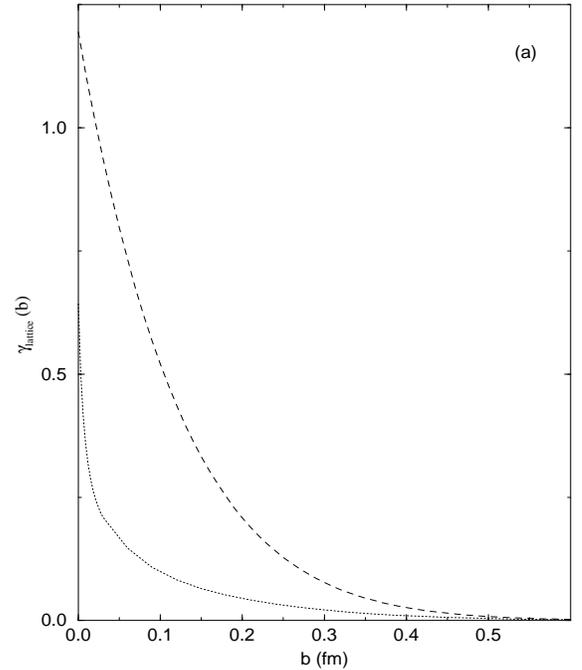,height=8.0cm,width=8cm}}
\vspace*{10.0mm}
\centerline{
\psfig{figure=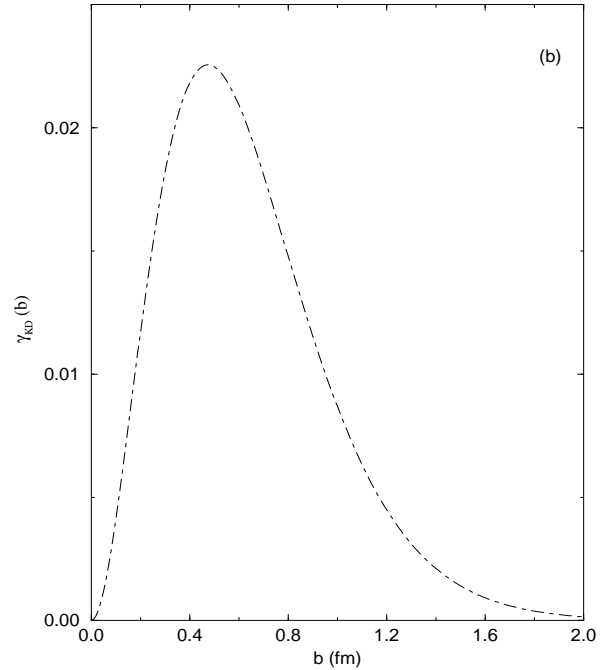,height=8.0cm,width=8cm}}
\vspace*{10.0mm}
\caption{Elementary (parton-parton) profile functions, Eq. 
(\protect\ref{perfilepsi}): (a) from lattice parametrizations with 
the divergent term (dotted), without the divergent term (dashed); 
(b) from the Kr\"{a}mer-Dosch ansatz.}
\label{gammab}
\end{figure}

At last, as before, the elementary profile is calculated through 
Eq. (\ref{perfilepsi}) now with normalization constant 

\begin{equation}
\eta={1\over 9(8\times 12)^2} .
\label{fator98dosch}
\end{equation}
By numerical integration \cite{nag93} we obtain 
the scattering amplitude. The results are shown in 
Figs.\ \ref{gammab} and \ref{amplq}.

\section{Discussion}
\label{sec:discu}

In this section we first sketch the conclusions coming directly 
from our calculations and then present discussions related to some 
phenomenological/semiempirical informations available on 
elementary amplitudes and also some theoretical issues.
\newpage

\subsection{Partial conclusions}
\label{subsec:partconc}

\begin{figure}\centerline{
\psfig{figure=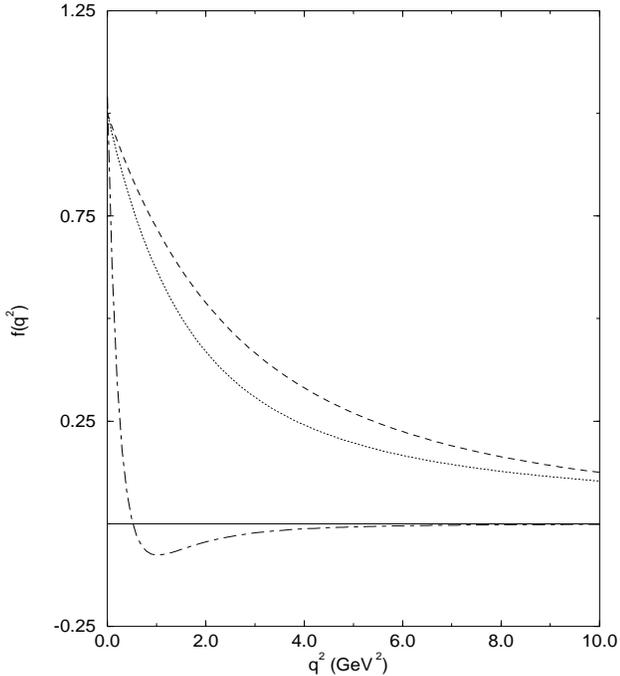,height=8.0cm,width=8cm}}
\vspace*{10.0mm}
\caption{Normalized elementary (parton-parton) scattering amplitudes, 
Eq. (\protect\ref{amplperfil}), from lattice parametrizations with 
the divergent term (dotted), without the divergent term (dashed), and 
from the Kr\"{a}mer-Dosch ansatz (dot-dashed).}
\label{amplq}
\end{figure}

The main results of our calculations are displayed in Figs. 
\ref{epsilonIepsilonII}, \ref{gammab} and \ref{amplq} (corresponding 
to eikonals, profiles, and amplitudes, respectively) and lead to the 
following conclusions.
\newpage

From Fig.\ \ref{epsilonIepsilonII}, the eikonal phase $\epsilon=
\epsilon_{\rm I}+\epsilon_{\rm II}$ from the recent lattice 
parametrizations (including the divergent term), has a maximum at 
$b=0$ and decreases smoothly as the impact parameter increases. 
Differently, the eikonal phase 
from the Kr\"{a}mer-Dosch ansatz presents the maximum at 
$b\approx 0.5$ fm and reaches zero at $b=0$. These distinct 
behaviors come from the differences between the correlators, 
shown in Fig. \ref{correldosch}.

Due to the structure of Eq. (\ref{perfilepsi}), the 
corresponding profiles present similar behaviors, as shown in Fig. 
\ref{gammab}. In this case we investigate the effect of the 
divergent term, referred to before. We see that, with or without 
the divergent term $1/|z|^4$ in the correlators $D$ and $D_1$, the 
profiles present maxima at $b=0$ and a smooth decrease with the 
impact parameter. With and without the divergence, the profiles 
reach $\sim 10\%$ of there maxima at $b=0.1$ and $0.2$ fm, 
respectively. We conclude that the term $1/|z|^4$ does not 
significantly alter the outputs leading only to a more concentrated 
and smaller profile near $b=0$.

As a consequence of the profiles, the corresponding 
scattering amplitudes also present analogous similarities and 
differences, as can be seen in Fig.\ \ref{amplq}. Normalization of 
the amplitudes to 1 at $q^2=0$, shows that both lattice 
amplitudes (with and without the divergent term), present a similar 
smooth decrease as the momentum transfer increases, going to zero 
through positive values. The effect of the divergent term is a 
faster decrease only. Differently, the amplitude from the 
Kr\"amer-Dosch ansatz decreases still faster, presenting a change of 
sign (zero) at $q^2\sim 0.5\;{\rm GeV}^2$ and goes asymptotically to 
zero through negative values. We see that according to our 
calculations, there is no agreement at all between profiles or 
amplitudes from lattice parametrizations and from the Kr\"amer-Dosh 
ansatz.

\subsection{Phenomenological and semiempirical informations}
\label{subsec:phenoempi}

Presently, some limited informations concerning elementary 
profiles and/or amplitudes are available from phenomelogical 
models and semiempirical analysis (model independent) as 
explained in what follows. The usual framework of analysis is 
the eikonal approximation in 
which the elastic {\it hadronic} amplitude $F$ is related to the 
eikonal function $\chi$ by \cite{glauber59}

\begin{equation}
F(q,s)=i\int bdb J_0(qb)\left\{1-\exp[i\chi(b,s)]\right\},
\label{amplhadtransf}
\end{equation}
where $\sqrt{s}$ is the center-of-mass energy. The hadronic 
amplitude is connected with physical observables as differential, 
total cross sections, etc., and the eikonal is defined in the 
context of each phenomenological model. In the simplest approach, 
represented by the Glauber or Chou-Yang formalism, the eikonal is 
expressed by \cite{menon93,czyz69,chou68}
\begin{equation}
\chi(b,s)=C\int qdqJ_0(qb)G_AG_Bf,
\label{eikoggf}
\end{equation}
where $G_A,\;G_B$ are hadronic form factors, $f$ the elementary 
(parton-parton) amplitude, and $C$ depends only on the energy.

{\it Phenomenology.} 
In the absence of theoretical predictions for both form factors and 
elementary amplitudes, models are yet characterized by different 
choices of parametrizations for these functions. As reviewed in Ref. 
\cite{menon93}, some models assume parametrizations for the 
{\it electromagnetic} form factors and others introduce {\it energy
-dependent} form factors. With quite different choices for the 
elementary amplitudes all these models present satisfactory 
descriptions of the experimental data. For this reason it is 
difficult to identify a ``phenomenological amplitude'' since it 
depends on the parametrizations used for the form factors and some 
other specific aspects of each model. So this puts serious 
limitations in attempts to compare theoretical - phenomenological 
results and we will return to this point in the next section.

However, even taking account of these limitations, it may be useful 
to see what kind of similarities could be estabilished. In this 
sense, based on the review of Ref. \cite{menon93} we could say 
the following.

\begin{figure}
\centerline{
\psfig{figure=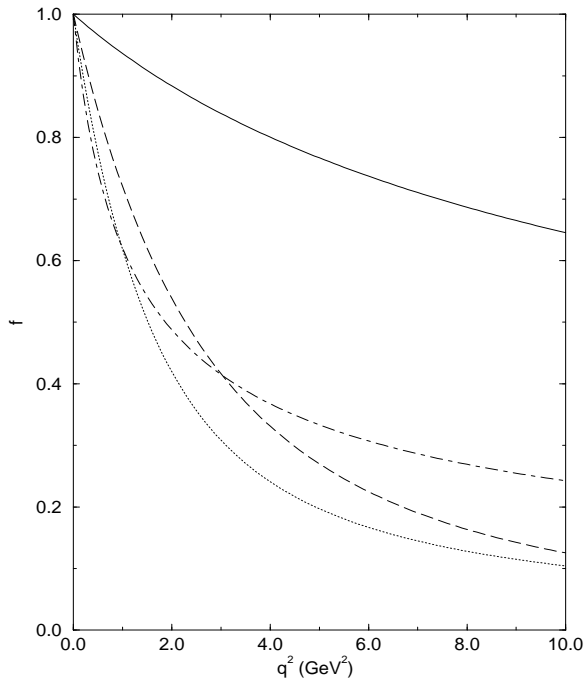,height=8.0cm,width=8cm}}
\vspace*{10.0mm}
\caption{Elementary amplitudes from lattice parametrizations, with 
(dotted) and without (dashed) the divergent terms (same as Fig.\ 
\protect\ref{amplq}), compared with the Glauber and Velasco 
parametrization, Eq. (\protect\ref{glaubervelasco}), for $pp$ at 
23 GeV (solid) and $\protect\overline{p}p$ at 546 GeV (dot-dashed).}
\label{figglaubvelas}
\end{figure}

(a) The behavior of both $\gamma(b)$ and $f(q)$ predicted 
from lattice parametrizations (Figs.\ \ref{gammab} and \ref{amplq}) 
are in qualitative agreement with the form introduced by Glauber and 
Velasco \cite{glauber84}

\begin{equation}
f_{\rm GV}(q,s)={1\over \sqrt{1+{q^2\over \alpha^2(s)}}} ,
\label{glaubervelasco}
\end{equation}
where $\alpha^2=7.14\;{\rm GeV}^2$ for $pp$ scattering at 
$\sqrt{s}=23$ GeV (with a phase factor in $f_{\rm GV}$) and 
$\alpha^2=0.625\;{\rm GeV}^2$ for $\overline{p}p$ at 546 GeV. 
The elementary amplitude decreases smoothly 
through positive values and goes asymptotically to zero, as shown in 
Fig.\ \ref{figglaubvelas}, for comparison with the results from 
lattice parametrizations.

(b) The profiles and amplitudes from the Kr\"{a}mer and 
Dosch ansatz (Figs.\ \ref{gammab} and \ref{amplq}) show 
qualitative agreement with the parametrizations used by Menon 
and Pimentel \cite{menon90}:

\begin{equation}
f_{\rm mBSW}={1-{q^2/\alpha^2}\over 1+{q^4/\alpha^4}},
\label{mBSW}
\end{equation}
where $\alpha^2=8.2\;{\rm GeV}^2$, independent of the energy or 
reaction, and the subscript means {\it modified} BSW (Bourrely, 
Soffer, and Wu), as explained in Ref. \cite{menon93}. The 
amplitude presents a change of sign and then goes asymptotically 
to zero through negative values. 

{\it Semiempirical analyses.} 
This last behavior is also predicted in the semiempirical analysis 
by Buenerd, Furget, and Valin \cite{furget90}. These authors used 
the fit to $pp$ differential cross section data introduced by Amaldi 
and Schubert \cite{amaldi80} and calculated the eikonal in the 
transfer momentum space. In the context of the Glauber--Chou-Yang 
model, Eq. (\ref{eikoggf}), this quantity reads

\begin{equation}
\chi(q,s)=CG_p^2f .
\label{eikoqg2f}
\end{equation}
The elementary amplitude $f(q)$ was then extracted at $\sqrt{s}=
23.5$ GeV by dividing $\chi(q,s)$ by parametrizations for the 
electromagnetic form factor (Felst parametrization and a dipole 
form used in the Bourrely-Soffer-Wu model \cite{furget90}) and 
normalizing the result at $q^2=0$ to one. The amplitude so 
extracted presents a change of sign at $q_0^2=8.6\;{\rm GeV}^2$, 
going to zero through negative values. In the CERN Intersecting 
Storage Ring (ISR) energy region 
($23.5\;{\rm GeV}\leq \sqrt{s}\leq 62.5\;{\rm GeV}$) the position 
of the zero decreases as the energy increases, reaching 
$q_0^2=5.0\;{\rm GeV}^2$ 
at $\sqrt{s}=62.5$ GeV \cite{furget90}.

However, this kind of procedure has two critical aspects.

(1) Since experimental data are available only in a limited 
interval of momentum transfer ($q^2 \leq 6.0\;{\rm GeV}^2$ in 
the above case) all kinds of extrapolations in the fits, allowed 
statistically, should be taken into account. This can be made by 
error propagation from the fits parameters, which, however, was 
not done in the above analysis.   

(2) The result for the amplitude 
yet depends on parametrizations for the form factor. The choice 
to use an electromagnetic form factor is an approximation since 
we are treating a hadronic and not an electromagnetic interaction. 
However, there is no theoretical or experimental information 
on hadronic matter form factors.

Recently, through fit procedures to $pp$ differential cross section 
data, Carvalho and Menon obtained the eikonal in the transfer 
momentum space, taking account of error propagation and also the 
effect of large momentum data \cite{carvalho97}. The result shows 
statistical evidence for a change of sign in the eikonal in the 
interval $5\;{\rm GeV}^2\alt q^2\alt 9\;{\rm GeV}^2$ at the ISR 
energy region \cite{carvalhome97} and asymptotical limit to zero 
through negative values. However, the movement of the zero with the 
energy cannot be inferred on statistical grounds. The analysis 
was performed through both a fit method and a numerical method. 
The former approach gives an average position of the zero at 
$q_0^2=7.1\pm 0.7\;{\rm GeV}^2$ and the later at $q_0^2=6.1\pm 
0.7\;{\rm GeV}^2$ \cite{carvalho97}. If, in the context of 
Eq. (\ref{eikoqg2f}), this zero is not associated with an hadronic 
form factor the result favors the amplitude from the 
Kr\"{a}mer-Dosch ansatz. Moreover, we found that the above 
positions of the zero are well reproduced if the gluonic 
correlation length in the Kr\"{a}mer-Dosch amplitude has the 
value $a\simeq 0.1$ fm. Specificaly, $a=0.0957$ fm for 
$q_0^2=7.1\;{\rm GeV}^2$ and $a=0.1040$ fm for $q_0^2=6.1\;
{\rm GeV}^2$. The result, in the case of the zero at $q_0^2=7.0\;
{\rm GeV}^2$, is shown in Fig.\ \ref{figmBSW}, together with the 
parametrization used by Menon and Pimentel for $\alpha^2=7.0\;
{\rm GeV}^2$ in Eq. (\ref{mBSW}).

\subsection{Theoretical and phenomenological issues}
\label{subsec:theolimi}

The physical picture in the Nachtmann approach is that of 
interaction of two long-lifetime partons with a common 
nonperturbative QCD vacuum, the partons traveling essentially 
on straight lightlike world lines. It is also assumed that the 
transversal momenta of the partons (related to the hadron's movement) 
are orders of magnitude smaller then the forward momenta. Therefore, 
this picture is characterized by {\it very small momentum transfer} 
and {\it asymptotically high-energy regime}. These are crucial 
aspects of the nonperturbative approach: They specify detailed 
frontiers, beyond which physical interpretations and/or comparisons 
with phenomenological information have no justification and this 
demands further discussion. In Figs.\ \ref{gammab} and \ref{amplq}, 
we display the results for the elementary profiles and amplitudes 
in a wide interval of impact parameter and momentum transfer. Based 
on the range of validity of the theoretical framework, we must 
see part of these curves as extrapolations to large momentum 
transfer and an essentially small impact parameter. Also, the 
elementary profiles and amplitudes do not depend on the energy 
since they are associated with the asymptotically high-energy 
limit. This leads to the following considerations.

\begin{figure}
\centerline{
\psfig{figure=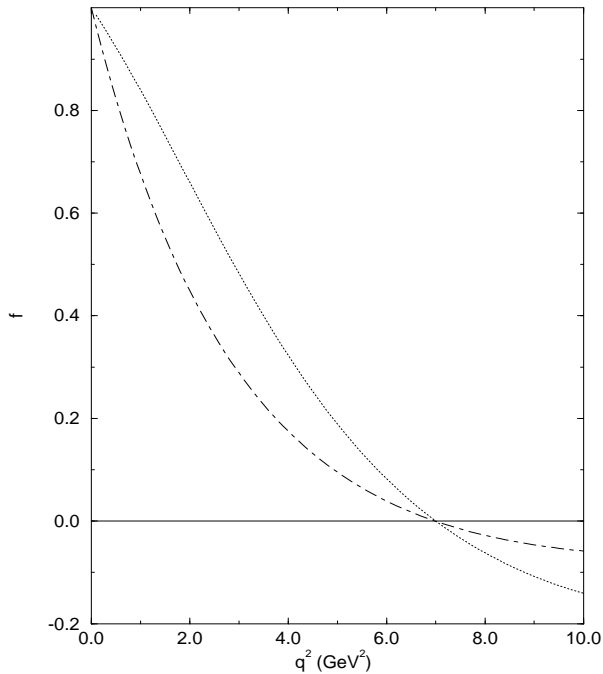,height=8.0cm,width=8cm}}
\vspace*{10.0mm}
\caption{Elementary amplitudes from Kr\"{a}mer-Dosch ansatz with 
correlation lenght $a=0.0957$ fm (dot-dashed) compared with the 
parametrization used by Menon and Pimentel, Eq. 
(\protect\ref{mBSW}), with $\alpha^2=7.0\;{\rm GeV}^2$ (dotted).}
\label{figmBSW}
\end{figure}

For the amplitudes obtained from the lattice 
parametrizations, the monotonical decrease to zero through 
positive values is clearly an extrapolation, at large transfer 
momenta. This behavior, however, could be broken, for example, by 
a specific model for that region, introducting scale properties, 
and even zeros for large $q^2$.    

The comparison displayed 
in Fig.\ \ref{figglaubvelas} between the lattice inspired 
amplitude and the Glauber-Velasco parametrization has also only 
a qualitative character, since in the last case the amplitude 
depends on the energy (and reaction). We see, however, a tendency 
to agreement as the energy increases, in the small transfer 
momentum region. 

The general agreement between the behaviors 
of the Kr\"{a}mer-Dosch amplitude and $f_{\rm mBSW}$ 
(Fig.\ \ref{figmBSW}) may have some more profound (but limited) 
meaning, since both amplitudes do not depend on the energy. Also, 
the connections with experimental data in the approaches of Refs. 
\cite{dosch94} and \cite{menon90} are obtained with energy 
dependences in the hadron's transverse wave function and hadronic 
form factors, respectively. This is very similar from a formal point 
of view: The energy dependence is ultimately associated with the 
hadronic radius. However, the limits in the transfer momentum 
region must be taken into account.

Another theoretical limitation that should be stressed concerns the 
correlation functions determined on the lattice. As referred to in 
Sec. \ref{subsec:resulatt} the theoretical results were obtained in 
the finite interval 0.1--1.0 fm and in this paper we used the 
parametrizations introduced in Ref. \cite{giacomo97}. With this 
assumption we found that the divergent term $1/|z|^4$ does not 
affect the profile or amplitudes in a substantial way 
(Figs.\ \ref{gammab}, \ref{amplq}). 
Now, what kind of information can we extract from the ``unknown'' 
region 0--0.1 fm? 
We observed that, neglecting the divergent term in the correlation 
function $D(z)$ by taking $B=0$ in Eq. (\ref{Ddegiacomo}), leads 
to a similar result as that obtained in the early calculation, 
Eq. (\ref{Ddedosch}). Specifically, in the former case the finite 
$D$ value at $|z|=0$ is $\approx 4.7\;{\rm GeV}^4$ and in the 
latter case, $\approx 5.0\;{\rm GeV}^4$. From Fig.\ \ref{correldosch} 
and Sec. \ref{subsec:kramedosch} we see that 
these values are yet near three times the finite value of the 
Kr\"{a}mer-Dosch correlator ($\approx 1.8\;{\rm GeV}^4$). This 
difference between the lattice extrapolations (without divergence) 
and Kr\"{a}mer-Dosch ansatz in the unacessible region of small 
distances (0.0--0.1 fm) has strong consequences, leading to the 
differences observed in the corresponding profiles 
(Fig.\ \ref{gammab}) and amplitudes (Fig.\ \ref{amplq}). That is, 
the existence of zero in the amplitude seems to be associated 
with the particular form of the correlator  at $0<|z|\alt 0.1$ fm 
(for example, exponential, Gaussian, etc.), or even at $0.1\leq |z|<
 0.4$ fm, and also to the value of the maximum at $|z|=0$. This 
small distance region, however, is not presently acessible to lattice 
calculations.

The results for the profiles and amplitudes showed a weaker 
dependence on the divergence of the type $|z|^{-4}$ than should be 
expected. Theoretically this kind of divergence may be associated 
either to operator product expansion (OPE) divergences or some 
specific high-energy gluon exchange process.

Our comparative analysis also brings some information on the 
gluonic correlation lenght {\it a} that should be discussed. This 
quantity is a very important parameter in the SVM applied to 
high-energy scattering. All the formalism is strongly sensitive to 
its value since, for example, total cross sections for 
parton-parton scattering is proportional to $a^{10}$ \cite{kramer90}. 
As explained in Sec. \ref{subsec:kramedosch}, the Kr\"amer-Dosch 
(KD) correlator function was determined by fit to early lattice 
results, from which they obtained $a=0.35$ fm and this value may 
reproduce parton-parton total cross sections of the order of $4$ 
mb \cite{kramer90}. On the other hand, from SU(2) lattice gauge 
theory, it was estimated $a = 0.1 \sim 0.2$ fm 
\cite{campostrini84}, and both early and recent parametrizations of 
the correlation functions from the lattice, Eqs. (\ref{Ddedosch}), 
(\ref{Ddegiacomo}), and (\ref{D1degiacomo}), lead to $a=0.22$ fm. 
In the context of the SVM, $a \sim 0.2$ fm  
means a very small contribution to parton-parton scattering in case 
of using KD ansatz, as discussed in \cite{kramer90}.

Concerning these distinct estimates, we showed in our analysis that 
for $a=0.35$ fm  the elementary amplitude for the Kr\"amer-Dosch 
correlator has a zero at $q^2 \simeq 0.5\;{\rm GeV}^2$ and that for 
$a\simeq 0.1$ fm  the position changes to $q^2 = 7.0\;{\rm GeV}^2$. 
Even taking account of the limitations related to the semiempirical 
analysis (Sec. \ref{subsec:phenoempi}), it is evident the existence 
of zeros at the  ISR energy region, in the interval $5.0  \leq q^2 
 \leq 9.0\;{\rm GeV}^2$ \cite{furget90,carvalhome97}. We conclude 
that, although this interval corresponds to finite energies, it 
seems to favor smaller values for the correlation lenght than 
0.35 fm.

At last it should also be stressed that the lattice results 
we used were obtained in the quenched approximation and 
therefore additional uncertainties may be considered. Recently, 
dynamical fermions have been taken into account in lattice 
calculations, allowing a full QCD treatment \cite{giacomo96}. 
However, since a reanalysis is in course, with improved statistics 
\cite{giacomopriv}, we will leave this subject for future 
consideration. It will also be suitable to compare these new 
results with those presented here.

\section{Final conclusions}
\label{sec:finconcl}

In the limit of extremely high energies (quarks on the 
light-cone) and small momentum transfer between partons, the 
nonperturbative approach connects the correlation functions with 
the parton-parton amplitudes. Making use of recent lattice 
parametrizations and also of the Kr\"amer-Dosch ansatz for these 
correlators, we calculated the corresponding elementary 
amplitudes in both impact parameter and transfer momentum spaces. 
In the case of lattice inputs the calculations were performed with 
and without the divergent term recently introduced.

We found that this divergence in the correlator does not 
substantially affect the normalized amplitudes and also that the 
results from lattice parametrizations and from the Kr\"amer-Dosch 
ansatz do not agree. Although the lattice parametrizations are 
statistically consistent with the theoretical points 
($\chi^2/N_{\rm DF}\sim 1.7$ \cite{giacomo97}) there is no real 
theoretical information in the interval 0.0--0.1 fm. This is a 
crucial point and our analysis furnished two novel results 
concerning possible connections between the short range behavior 
of the correlators and the final results for the elementary 
amplitudes. On the one hand, Fig.\ \ref{amplq} shows that the 
divergent term $1/|z|^4$ causes only a faster decrease of the 
amplitude as the momentum transfer increases, that is, it does not 
generate local minima or zeros. The amplitude coming from this 
particular lattice parametrization, with or without the divergent 
term, is characterized by a monotonical decrease with the 
momentum transfer through positive values. On the other hand, in 
the case where the correlator may be finite at $|z| = 0$, its 
maximum value and also its shape at small $|z|$ ($|z| 
\lesssim 0.4$ fm) strongly affect the behavior of the amplitude, 
as shown in Fig.\ \ref{amplq} and quantitatively discussed in 
Sec. \ref{subsec:theolimi}: Zeros (change of sign) and local 
minima may be generated. That is, despite the present lack of 
theoretical information in the short range $0.0 - 0.1$ fm, we 
can infer, in the context of the stochastic vacuum model, what 
kind of behavior should be expected for the elementary amplitude.

Attempts to compare these results with phenomenological and/or
semiempirical information on the amplitudes are limited. The 
reason is that this information comes from an analysis at finite 
energies (presently available) and the amplitudes depend on 
particular model assumptions or specific parametrizations for 
the form factors.

On the other hand, as referred to in Sec. \ref{sec:discu}, the 
distinct results for the amplitudes from the lattice and 
Kr\"amer-Dosch ansatz show qualitative agreement with different 
model predictions such as Glauber-Velasco and Menon-Pimentel, 
respectively. Furthermore, semiempirical analysis could favor the 
Kr\"amer-Dosch amplitude, but with a gluonic correlator length 
smaller than 0.35 fm.

We conclude that, even taking account of all the limitations 
referred to (Secs. \ref{subsec:phenoempi} and 
\ref{subsec:theolimi}), these agreements suggest 
possible links between theory and phenomenological or 
semiempirical analyses. Further investigations along these lines 
may be important as a source of feedback for theoretical 
(nonperturbative) ideas, meanly concerning suitable considerations 
on energy dependences and small distance phenomena.

\acknowledgments

We are grateful to Professor A. Di Giacomo for a critical reading 
of the manuscript and valuable discussions and suggestions. We give 
thanks to Capes and CNPq for financial support.


\begin{references}
\bibitem{menon93}M. J. Menon, Phys. Rev. D {\bf 48}, 2007 (1993); 
A. F. Martini, M. J. Menon, and D. S. Thober, {\it ibid.} 
{\bf 54}, 2385 (1996); in {\it Frontiers in Strong Interactions, 
Proceedings of the VIth Blois Workshop on Elastic and Diffractive 
Scattering}, edited by P. Chiappetta, M. Haguenauer, and J. 
Tr\^{a}n Thanh V\^{a}n (Editions Fronti\'{e}res, Gif-sur-Yvette, 1996) 
p. 121.
\bibitem{furget90}C. Furget, M. Buenerd, and P. Valin, Z. Phys. C
{\bf 47}, 377 (1990).
\bibitem{landshoff87}P. V. Landshoff and O. Nachtmann, Z. Phys. C
{\bf 35}, 405 (1987).
\bibitem{nachtmann91}O. Nachtmann, Ann. Phys. (N.Y.) {\bf 209}, 436 
(1991).
\bibitem{kramer90}A. Kr\"{a}mer and H. G. Dosch, Phys. Lett. B 
{\bf 252}, 669 (1990).
\bibitem{dosch87}H. G. Dosch, Phys. Lett. B {\bf 190}, 177 (1987).
\bibitem{dosch88}H. G. Dosch and Yu. A. Simonov, Phys. Lett. B
{\bf 205}, 339 (1988).
\bibitem{kampen74}N. G. van Kampen, Physica (Amsterdam) {\bf 74}, 
215 (1974).
\bibitem{simonov89}Yu. A. Simonov, Nucl. Phys. {\bf B204}, 67 (1989).  
\bibitem{dosch94}H. G. Dosch, E. Ferreira, and A. Kr\"{a}mer, 
Phys. Rev. D {\bf 50}, 1992 (1994).
\bibitem{giacomo92}A. Di Giacomo and H. Panagopoulos, 
Phys. Lett. B {\bf 285}, 133 (1992).
\bibitem{giacomo97}A. Di Giacomo, E. Meggiolaro, and H. 
Panagopoulos, Nucl. Phys. {\bf B483}, 371 (1997).
\bibitem{grandel95}U. Grandel and W. Weise, Phys. Lett. B {\bf 356}, 
567 (1995).
\bibitem{campostrini89}M. Campostrini, A. Di Giacomo, M. Maggiore, 
H. Panagopoulos, and E. Vicari, Phys. Lett. B {\bf 225}, 403 (1989).
\bibitem{giacomo90}A. Di Giacomo, M. Maggiore, and S. Olejnik,
Phys. Lett. B {\bf 236}, 199 (1990); Nucl. Phys. {\bf B347}, 441 
(1990).
\bibitem{michael88}C. Michael and M. Teper, Nucl. Phys. {\bf B305}, 
453 (1988).
\bibitem{nag93}NAG FORTRAN Library Manual, Mark 16, 1993; S. 
Wolfram, MATHEMATICA (Addison-Wesley, New York, 1991).
\bibitem{james92}F. James and M. Roos, MINUIT--Function 
Minimization and Error Analysis, CERN Report No. D506 (CERN, 
Geneva, 1992).
\bibitem{glauber59}R. J. Glauber, in {\it Lectures in Theoretical 
Physics}, edited by W. E. Britten {\it et al.} (Interscience, New 
York, 1959), Vol. I, p. 315.
\bibitem{czyz69}W. Czy\.z and L. C. Maximon, Ann. Phys. (N.Y.) 
{\bf 52}, 59 (1969); V. Franco and G. K. Varma, Phys. Rev. C
{\bf 18}, 349 (1978).
\bibitem{chou68}T.T. Chou and C.N. Yang, Phys. Rev. {\bf 175}, 1832 
(1968).
\bibitem{glauber84}R.J. Glauber and J. Velasco, Phys. Lett. B 
{\bf 147}, 380 (1984); in {\it Proceedings of the Second 
International Conference on Elastic and Diffractive Scattering}, 
New York, 1987, edited by K. Goulianos (Editions Fronti\`eres, 
Gif-sur-Yvette, 1988), p. 219.
\bibitem{menon90}M.J. Menon and B.M. Pimentel, Hadronic J. 
{\bf 13}, 325 (1990); M.J. Menon, in {\it Proceedings of the 
Fourth International Conference on Elastic and Diffractive 
Scattering}, Elba, Italy, 1991, edited by P. Cervelli and 
S. Zucchelli [Nucl. Phys. B (Proc. Suppl.) 
{\bf 25}, 94 (1992)]; Canadian J. Phys. {\bf 74}, 594 (1996).
\bibitem{amaldi80}U. Amaldi and K.R. Schubert, Nucl. Phys. 
{\bf B166}, 301 (1980).
\bibitem{carvalho97}P. A. S. Carvalho and M. J. Menon, in {\it 
VIII Workshop on Hadronic Interactions}, edited by Y. Hama, F. 
S. Navarra, and M. Nielsen (Instituto de F\'{\i}sica - USP, 
S\~ao Paulo, 1997), p. 12. 
\bibitem{carvalhome97}P. A. S. Carvalho and M. J. Menon, Phys. 
Rev. D {\bf 56}, 7321 (1997).
\bibitem{campostrini84}M. Campostrini, A. Di Giacomo, and G. 
Mussardo, Z. Phys. C {\bf 25}, 173 (1984).
\bibitem{giacomo96}A. Di Giacomo, E. Meggiolaro,
and H. Panagopoulos, Phys. Lett. B {\bf 408}, 315 (1997).
\bibitem{giacomopriv}A. Di Giacomo (private communication).

\end{references}
\end{document}